\documentclass[journal,twoside,web]{ieeecolor}
\usepackage{generic}
\usepackage{amsmath,amssymb,amsfonts}
\usepackage{algorithmic}
\usepackage{graphicx}
\usepackage{textcomp}

\usepackage[utf8]{inputenc}
\usepackage[T1]{fontenc}
\usepackage[english]{babel}
\usepackage[hidelinks, unicode]{hyperref}
\usepackage[table,xcdraw]{xcolor}
\usepackage{url}
\usepackage{csquotes}
\usepackage{subcaption}
\usepackage[style=ieee, backend=biber, url=false, doi=true, eprint=false, isbn=false, date=year, 
]{biblatex}

\usepackage{xfrac}
\usepackage{nidanfloat}
\usepackage{orcidlink}

\AtEveryBibitem{%
  \clearfield{note}%
}

\graphicspath{ {images/} }

\def\articletitle{Closed loop BCI System for Cybathlon 2020}

\def\BibTeX{{\rm B\kern-.05em{\sc i\kern-.025em b}\kern-.08em
    T\kern-.1667em\lower.7ex\hbox{E}\kern-.125emX}}
\begin{document}

\title{\articletitle}
\author{
% Csaba Köllőd, \IEEEmembership{Member, IEEE}, András Adolf, \IEEEmembership{Member, IEEE}, Gergely Márton, \IEEEmembership{Fellow, IEEE}, Moutz~Wahdow, \IEEEmembership{Member, IEEE}, Ward Fadel \IEEEmembership{Member, IEEE} and István Ulbert \IEEEmembership{Fellow, IEEE}
Csaba Köllőd \orcidlink{0000-0003-3817-6709}, András Adolf \orcidlink{0000-0001-7855-1492}, Gergely Márton \orcidlink{0000-0003-4359-0111}, Moutz~Wahdow \orcidlink{0000-0003-4566-8357}, Ward Fadel and István Ulbert \orcidlink{0000-0001-9941-9159}

\thanks{\textbf{Submission date:} 2022.04.28. This work was supported by the Ministry of Innovation and Technology of Hungary under Grant KDP-2021-12 1020483.}
\thanks{Csaba Köllőd is with the Roska Tamás Doctoral School of Sciences and Technology, Budapest, Hungary (e-mail: kollod.csaba@itk.ppke.hu).}
\thanks{András Adolf is with the Roska Tamás Doctoral School of Sciences and Technology, Budapest, Hungary (e-mail: adolf.andras@itk.ppke.hu).}
\thanks{Gergely Márton is with the Institute of Cognitive Neuroscience and Psychology, Research Centre for Natural Sciences, Budapest, Hungary (e-mail: marton.gergely@ttk.hu).}
\thanks{Moutz~Wahdow is with the Roska Tamás Doctoral School of Sciences and Technology, Budapest, Hungary (e-mail: wahdow.moutz@itk.ppke.hu).}
\thanks{Ward Fadel is with the Roska Tamás Doctoral School of Sciences and Technology, Budapest, Hungary (e-mail: fadel.ward@itk.ppke.hu).}
\thanks{István Ulbert is with the Roska Tamás Doctoral School of Sciences and Technology, Budapest, Hungary and with the Institute of Cognitive Neuroscience and Psychology, Research Centre for Natural Sciences, Budapest, Hungary (e-mail: ulbert.istvan@ttk.hu).}
}

\maketitle

\begin{abstract}
We present our Brain-Computer Interface (BCI) System, developed for the BCI discipline of Cybathlon 2020 competition. In the BCI discipline, subjects with tetraplegia are required to control a computer game with mental commands. The absolute of the Fast-Fourier Transformation amplitude was calculated as a feature (FFTabs) from one-second-long electroencephalographic (EEG) signals. To extract the final features, we introduced two methods, namely the Feature Average, where the average of the FFTabs for a specific frequency band was calculated, and the Feature Range, which was based on generating multiple Feature Averages for non-overlapping 2 Hz wide frequency bins. The resulting features were fed to a Support Vector Machine classifier. The algorithms were tested on the PhysioNet database and our dataset, which contains 16 offline experiments recorded with 2 tetraplegic subjects. 27 gameplay trials (out of 59) with our tetraplegic subjects reached the 240-second qualification time limit. The Feature Average of canonical frequency bands (alpha, beta, gamma, and theta) were compared with our suggested range30 and range40 methods. On the PhysioNet dataset, the range40 method combined with an ensemble SVM classifier significantly reached the highest accuracy level (0.4607), with a 4-class classification, and outperformed the state-of-the-art EEGNet.
\end{abstract}

\begin{IEEEkeywords}
% Enter key words or phrases in alphabetical 
% order, separated by commas. For a list of suggested keywords, send a blank 
% e-mail to keywords@ieee.org or visit \underline
% {http://www.ieee.org/organizations/pubs/ani\_prod/\discretionary{}{}{}keywrd98.txt}
BCI, EEG, SVM, FFT, Normalization
\end{IEEEkeywords}

\section{Introduction}
Cybathlon Competition was originally introduced on 5\textsuperscript{th} October 2014 \cite{riener_cybathlon_2014}. This competition is also called the “Bionic Olympics”,where research groups, industrial companies, and technology providers can present their products, applications, and technologies in six disciplines with the help of disabled subjects, called pilots in Cybathlon terminology. The six disciplines are Brain-Computer Interfaces (BCI), Functional Electrical (FES) Bike race, Leg Prosthesis, Powered Arm Prosthesis, Powered Exoskeleton, and Powered Wheelchair. 

Brain-Computer Interfaces are integrated systems that include software and hardware components. As it is presented by Wolpaw et al.\ \cite{wolpaw_braincomputer_2002} these systems record bioelectrical signals from the brain and convert them to computer commands. In the BCI discipline of Cybathon, pilots with quadriplegia compete to reach the finish line in a car-racing-like computer game. Each pilot must control their avatar using well-timed imagined mental commands recorded by Electroencephalography (EEG). The recorded raw EEG data are often affected by internal or external noises, such as eye blinking, swallowing, electric powerline noise, or motion artifacts. Using any artifact for control is strictly forbidden, and the implementation of a filtering and artifact rejection algorithm is required. The computer game can be controlled with 3 active commands plus the absence of any commands. The pilots are required to reach the finish line in under 240 seconds. 

During the FES (Functional Electrical Stimulation) discipline, pilots with paraplegia have a tricycle competition, where the muscle activities are generated using functional electrical stimulation. Under the Leg and Arm Prosthesis disciplines, obstacle courses should be completed which are designed to highlight the capabilities and usability of the prosthesis in different life situations. The exoskeleton helps paraplegic pilots to stand up, walk and climb stairs, while the Powered wheelchair discipline challenges the pilots to navigate through stairs, different types of roads, and crowds. This article focuses on the BCI discipline.

Perdikis et al.\ \cite{perdikis_cybathlon_2018} participated in the first Cybathlon competition with two pilots from a team called Brain Tweakers. They detected artifacts using electrooculographic (EOG) electrodes and the FORCe algorithm. They conducted Laplacian derivation on pure EEG signals as special filtering, followed by Power Spectral Density calculation on 2 Hz wide frequency intervals. The created features were classified with the Gaussian mixture model. They used two class motor imaginary signals, namely both hands and both legs movements. In order to meet the requirements of controlling the game, they implemented a strategy: If two different types of commands were generated under a configurable width time window, the third active control signal was sent to the game. Instead of further developing and finetuning the control algorithm, they focused on the pilots' training, claiming that learning to purposefully modulate brain waves also dramatically impacts the BCI system's usability. As proof of their hypothesis, one of their pilots won the Cybathlon 2016 competition, and they successfully used their algorithm even in  Cybathlon BCI Series 2019 and Cybathlon Global Edition 2020 competition \cite{tortora_neural_2022} as WHI Team, reaching first place.

Team MIRAGE91 from Gratz \cite{statthaler_cybathlon_2017} created an online artifact detection system that included a blinking detector by thresholding on the band power of the AFz electrode and autoregressive modeling to explore high deviations. For feature extraction, the Common Spatial Pattern was used with shrinkage regularized Linear Discriminant Analysis for classification at Cybathlon 2016. They implemented a 3-class paradigm with a thresholding strategy with the following motor imagery (MI) tasks: Left Hand, Right Hand, and Both Feet. The control output was only sent to the game if the classification probability met the threshold. They could show a smooth learning curve with their pilots; however, they faced unexpected issues and performed way beyond their training results under the competition.     For the Cybathlon BCI Series 2019 and Cybathlon Global Edition 2020 competition, they improved their algorithm, including a novel adaptive thresholding algorithm \cite{hehenberger_long-term_2021} for controlling the output of the BCI.

The Nitro1 team \cite{benaroch_long-term_2021} was focusing on minimizing the within- and between-session variability and shifts with the aid of the Riemann framework. The generated features were projected to a common reference before making a classification with a Minimum Distance to Mean classifier. 4 class BCI system were created with 2 MI classes (Left Hand, Right Hand), mental subtraction and the idle state. A blinking detector and thresholding technique on the absolute of the EEG signals were used to reject features with artifacts. 

Team SEC FHT \cite{robinson_design_2021} implemented an artifact removal algorithm, which detected EOG artifacts with Pearson’s correlation and interpolated the affected channels. On the purified EEG data Filter Bank Common Spatial Pattern was calculated and Gaussian kernelled Support Vector Machine (SVM) was used to classify the features. To evoke control signals 4 MI tasks were used (Left Hand, Right Hand, Both Legs, Rest). The team investigated the precision of the control system, with respect to the training protocol. In their analysis, they compared the offline arrow based, the offline game based, and the online gameplay training. They showed that the best performance was achieved with training data derived from the online gameplay where the pilot received immediate feedback about the correctness of the executed commands.

In parallel with the Cybathlon, many efforts were concentrated on developing suitable offline BCI systems that properly classify EEG signals stemming from MI tasks. \cite{ang_filter_2012, lawhern_eegnet_2018, riyad_mi-eegnet_2021, zhang_eeg-based_2017} 
Many of these systems use artificial neural networks as a classifier \cite{fadel_chessboard_2020,fadel_multi-class_2020, riyad_mi-eegnet_2021, roots_fusion_2020}. 
The EEGNet \cite{lawhern_eegnet_2018}, created by Lawhern et al., is one of the state of the art networks. Compared to simple classification methods, one advantage of this algorithm is that it does not require any feature-extracted signal. Instead, it only requires raw EEG data in a matrix form and learns features similar to Filter Bank Common Spatial Pattern \cite{ang_filter_2012}. However, simple classifier methods, such as K-Nearest-Neighbor \cite{gedik_classification_2021, varsehi_eeg_2021}, Linear Discriminant Analysis \cite{gedik_classification_2021, gwon_alpha_2021} or Support Vector Machine (SVM) \cite{huang_multi-view_2022, jin_correlation-based_2019, li_multi-scale_2020} are also preferred in BCI systems, where computational requirements are planned as modest, as in our approach. Therefore, we selected SVM for our algorithm and compared it with the state-of-the-art EEGNet. Most of the comparison investigations of the classifiers use one of the BCI Competition datasets \cite{blankertz_bci_2004, blankertz_bci_2006, sajda_data_2003, tangermann_review_2012}. However, these datasets contain only a few numbers of subjects $(\leq10)$. In \cite{fan_bilinear_2020, roots_fusion_2020, varsehi_eeg_2021} is used. The participation of 109 subjects created this database; therefore, we selected it for comparison to make it statistically significant.

Our paper aims to present the complete development path of the BCI system, which was created from scratch for the BCI competition of Cybathlon Global Edition 2020 (Cybathlon 2020) by the Hungarian research group team called Ebrainers. The development started by creating signal processing and classification methods. Algorithms were tested offline on the PhysioNet dataset. In parallel, we also involved tetraplegic subjects in recording experiments with them. After evaluating the recorded data and algorithm, our real-time working BCI system was created to control the video game, which the organizers of Cybathlon 2020 provided. Regular experiments were conducted till 5\textsuperscript{th} March 2020. Due to the pandemic, the experiments and our participation in Cybathlon 2020 were canceled.

\section{Materials and methods}
To develop a BCI system, which can be used to control a computer game, the first step is to design a feature extraction and a classification method and test them on a reliable dataset. We aimed to create a subject-specific BCI instead of a general one, as was reported in \cite{bria_sinc-based_2021}, and \cite{lawhern_eegnet_2018}, to be superior in classification results. 

The development required a reliable database, which included not even a large amount of EEG signals but also the correct trigger points of the experimental tasks with appropriate labels as annotations. Accurate labeling is essential for appropriately testing the precision of classification and other needed algorithms.

\subsection{Physionet database}
The EEG Motor Movement/Imagery Dataset, available on PhysioNet (Physionet) \cite{goldberger_ary_l_physiobank_2000}, is one of the largest databases based on MI tasks, for which the EEG signals were acquired by the BCI2000 system \cite{schalk_bci2000_2004}. The Physionet dataset contains EEG records from 109 subjects, which were recorded with a 64 channeled 10-20 EEG system.

In brief, the experimental paradigm of Physionet was built as follows: At the beginning of the experiment, each subject had two one-minute-long baseline sessions, where they had to sit calm, first eyes open, then closed. After this session, the movement execution and imagination period started. First, the subjects had to execute overtly left-hand and right-hand movements, which was followed by an imaginary session with covert movement (no voluntary movement). Then the executed and imagined sessions involving both hands and feet were followed. These tasks were repeated 3 times, which resulted in 14 sessions in an experiment, in addition to the two baseline sessions. Each executed and imagined movement took 4 seconds, followed by a 4-second-long resting period. 

For some minor problems in the data acquisition of the Physionet database, we decided to exclude subjects 88, 89, 92, and 100 for the following reasons. In the case of subject 89, we found that the labels are incorrect. The records of subjects 88, 92, and 100 differ from the main description of the database. The execution of the task and the resting phase were changed to 5.125 and 1.375 seconds, respectively. Moreover, a 128 Hz sampling frequency was used instead of the original 160 Hz. These problems were reported in \cite{fan_bilinear_2020, roots_fusion_2020}.

\subsection{Signal Processing and Classification}
\begin{figure*}[htp]
  \centering
  \includegraphics[width=\textwidth]{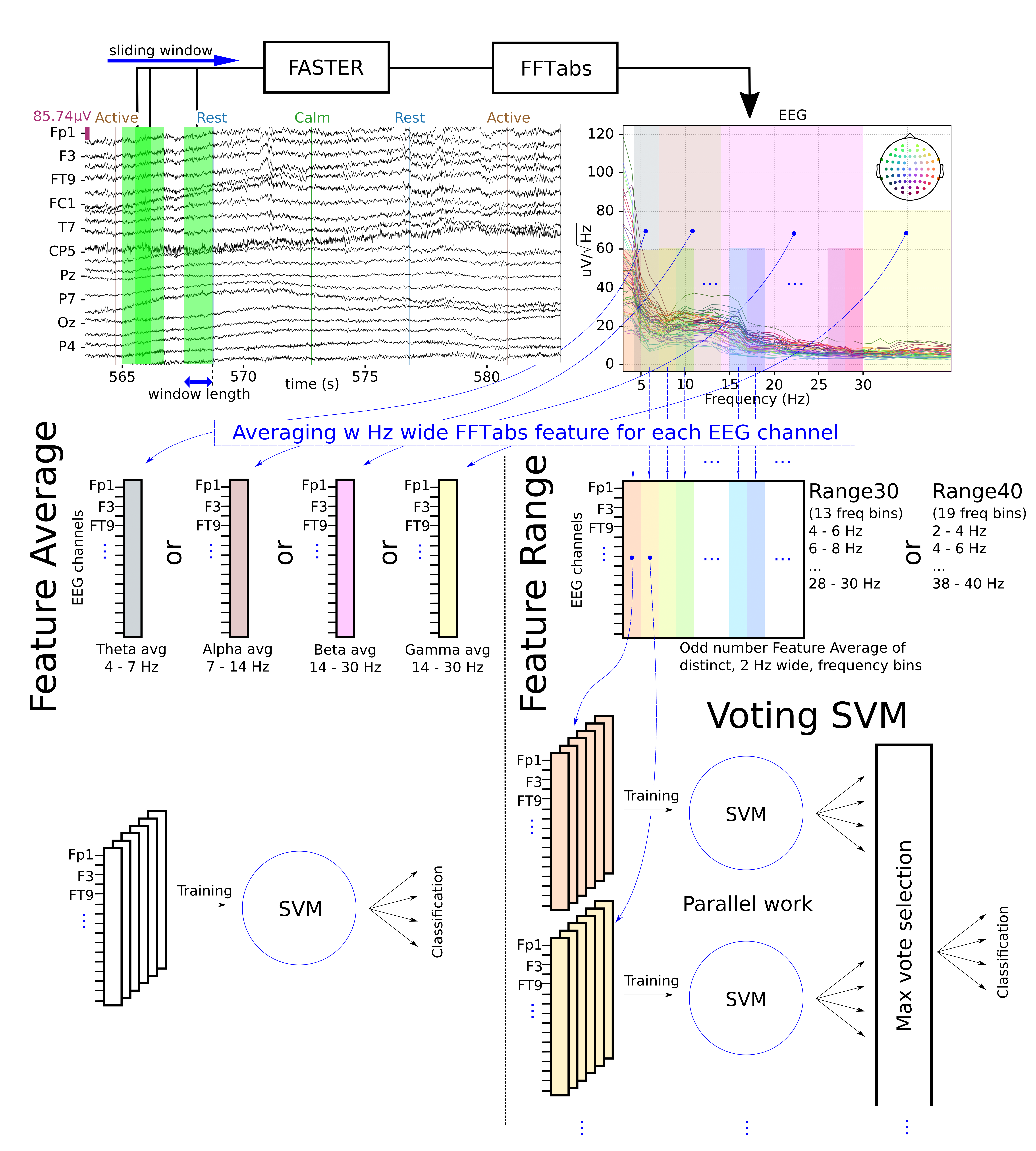}
  \caption{\textbf{BCI pipeline} -- FASTER  algorithm was used to remove EOG and EMG artifacts from raw EEG signals. From a 1-second long window, the absolute of the Fast Fourier Transformation (FFTabs) was calculated as a feature for each EEG channel. In the case of the Feature Average method, the average of the FFTabs was calculated for each channel in a given frequency range which could be one of the canonical EEG bands (Theta, Alpha, Beta, Gamma). These features were used to train a Support Vector Machine (SVM). The Feature Range method uses and extends the Feature Average. The Feature Average was calculated for an odd number of distinct, 2 Hz wide frequency bins. Each feature, corresponding to a frequency bin, was used to train a separate SVM, and the final classification result was determined as the max vote of all SVM units.}
  \label{fig:sig-proc}
\end{figure*}

In this section, the core part of the BCI system is presented. Python programming language, an open-source, cross-platform language with many libraries, including state-of-the-art EEG signal processing and machine learning packages such as MNE \cite{gramfort_meg_2013} and TensorFlow \cite{abadi_tensorflow_2016}, was selected to design and develop our BCI system. It includes artifact rejection, feature extraction, and classification methods. The complete signal processing architecture is presented in Figure \ref{fig:sig-proc} and available at: \url{https://github.com/kolcs/bionic_apps}

\subsubsection{Artifact rejection algorithm}
The Fully Automated Statistical Thresholding algorithm (FASTER), published by Nolan et al. \cite{nolan_faster_2010}(Nolan et al., 2010), was used for artifact rejection. The source of our python implementation was developed by Vliet \cite{vliet_wmvanvlietmne-faster_2021}. This algorithm was designed for offline brain signal processing; however, we aimed also to use it in our real-time BCI system. To achieve this goal, an online version of this algorithm was implemented. The method starts with the offline processing of the prerecorded training dataset of the Online Paradigm. During the process, the artifact-rejector saves parameters detailed below.

The offline algorithm contains four steps designed to discard channels, epochs, and components where the Z-score of given parameters (detailed below) is greater than 3.
As the first step, it removes those EEG channels where the Z-score of the variance, the correlation, or the Hurst exponent is over the limit. These bad channels are saved and given later to the online process.
In the next step, epochs are discarded concerning the amplitude range, deviation from the channel average, and variance parameters. This step is omitted during the online algorithm, as we need all epochs to classify.

The third step aims to remove artifact-related components of the signal. It utilizes independent component analysis (ICA), separating time-dependent data into statistically independent waveforms. The algorithm outputs the mixing matrix, with which the multiplication transforms EEG data into Independent Components (IC). The FastICA implementation of the Scikit-learn package \cite{pedregosa_scikit-learn_2011} was used to perform this computation. The first parameter on which a component is omitted is the correlation with the electrodes closest to the eyes. This way, the blinking artifacts are aimed to be filtered out. Those ICs are also discarded where the Z-score of kurtosis, power gradient, Hurst exponent, or median gradient is greater than 3. To transform the ICs back to the time domain, multiplication by the inverse of the mixing matrix is performed.

In the fourth step, bad channels in individual epochs are detected. The examined parameters are variance, median gradient, amplitude range, and channel deviation. All channels labeled as bad are interpolated with spherical spline interpolation. This step remains as it is for the online algorithm.
Although the original FASTER algorithm consists of 5 steps, the last one, detecting artifacts across subjects, was omitted to create a subject-specific process. 

To summarize, our online FASTER algorithm gets globally bad channels and the ICA matrix as initial parameters. During online processing, it performs component filtering and interpolates the incoming epochs’ globally and individually bad channels.

\subsubsection{Feature extraction}
During the feature extraction, epochs were created from the raw signal concerning the trigger points. One epoch started from the trigger point and ended 4 seconds later, as the experimental subjects executed MI tasks during this period. From the created epochs, shorter, 1-second-long EEG windows were cut with a 0.1-second shift to create a sliding window behavior.

Each EEG window calculated the absolute of the complex Fast Fourier Transformation \cite{cooley_algorithm_1965} (FFTabs) for each EEG channel as a frequency domain feature. The following methods used the FFTabs function.

\paragraph{Feature Average}
In our	Feature Average method, the numerical average of the FFTabs was calculated in a specified frequency range for each EEG channel. The following equation represents this:

\begin{equation}
    feature_{ch_i}=\frac{1}{N}\sum_{f=f_{min}}^{f_{max}}{{\mathrm{FFTabs}}_{ch_i}\left(f\right)}
\end{equation}
where $ch_i$ is the $i^{th}$ EEG channel and $N$ is the number of FFTabs samples in the defined $\left[f_{min},f_{max}\right]$ frequency range. 

This process can be interpreted as cropping the FFTabs to the selected frequency interval and squeezing this matrix to a channel\_number x 1 feature vector. The edges of the frequency range are parameters that can be selected corresponding to one of the canonical EEG bands: theta (4-7 Hz), alpha (7-14 Hz), beta (14-30 Hz), and gamma (30-40 Hz).

\paragraph{Feature Range}
The Feature Range method uses and extends the Feature Average. We aimed to enhance the information content of the calculation compared to the Feature Average. This method generated multiple Feature Averages for non-overlapping, 2 Hz wide frequency bins. This method has two parameters defining the lowest and highest frequency edge. The first frequency bin ranges from $f_{low}$ to $f_{low}+2Hz$ and the last from  $f_{high}-2Hz$ and $f_{high}$. We created two feature sets from Feature Range, called range30 and range40, where the numbers correspond to $f_{high}$. In the case of range30 $f_{low}=4Hz$ and $f_{high}=30Hz$, which resulted in 13 frequency bins in total. Therefore, the size of the feature matrix is channel number x 13. In the case of range40 $f_{low}=2Hz$ and $f_{high}=40Hz$, which resulted in 19 frequency bins in total.

\subsubsection{Feature normalization}
After calculating the features, L2 normalization was conducted after calculating the features to improve the classification results, as reported in \cite{raju_study_2020}. The L2 normalization is defined as follows:
\begin{equation}
     X_{l2}=X \Big/ \sqrt{\sum_{k=1}^{n}\left|x_k\right|^2}
\end{equation}

\subsubsection{Support-Vector Machine based classifiers}
As a classification tool, we used the Support Vector Machine method, as it has modest computational demands on our system, and also frequently used in BCI applications \cite{huang_multi-view_2022, jin_correlation-based_2019, li_multi-scale_2020, robinson_design_2021}. 

The original problem of Support Vector Machine (SVM) is formulated by Vapnik \cite{boser_training_1992}. If there is a given training set $\left(\mathbf{x}_i,y_i\right),\ i=1,\ldots,k$ where $\mathbf{x}_i\in R^n$ is a training sample with $y_i\in\{-1,1\}$ label the SVM has to solve the following optimization problem:

\begin{equation}
    \min_{\mathbf{w}, b,
    \mathbf{\xi}}{\frac{1}{2}}\mathbf{w}^T\mathbf{w}+C\sum_{i=1}^{k}\xi_i
\end{equation}
subject to

\begin{equation}
    y_i\left(\mathbf{w}^T\phi\left(\mathbf{x}_i\right)+b\right)\geq1-\xi_i,\quad\xi_i\geq0
\end{equation}
where $\phi$ is a nonlinear function which can map $x_i$ to a higher dimensional feature space and $C>0$ is a penalty hyperparameter of error term. The $K\left(\mathbf{x}_i,\mathbf{x}_j\right)\equiv\phi\left(\mathbf{x}_i\right)^T\phi\left(\mathbf{x}_j\right)$ term is called as the kernel function, which can be an arbitrary mathematical equation. The most frequently used ones are the following:
\begin{itemize}
    \item linear: $K\left(\mathbf{x}_i,\mathbf{x}_j\right)=\mathbf{x}_i^T\mathbf{x}_j$
    \item polynomial: $K\left(\mathbf{x}_i,\mathbf{x}_j\right)=\left(\gamma{\mathbf{x}_i}^T\mathbf{x}_j+r\right)^d,\quad \gamma>0$
    \item Radial-basis-function (RBF):\\ $K\left(\mathbf{x}_i,\mathbf{x}_j\right)=exp\left(-\gamma \|\mathbf{x}_i-\mathbf{x}_j\|^2\right),\quad \gamma>0$ 
\end{itemize}
where $\gamma$, $r$ and $d$ are the kernel parameters, which also can be considered as hyperparameters.

For solving the SVM problem, we used the Scikit-learn package (Pedregosa et al., 2011), which includes many different, efficient implementations and other useful machine-learning tools. From the given SVM classifiers, in Scikit-learn, the SVC class was selected, which defines an RBF kernelled Support Vector Classifier. The default hyperparameters were used in all experiments for all classifications. In the case of the Feature Average method, the channel number x 1 size feature vectors were used to train and classify the data.  

In the case of the Feature Range method, the feature vectors of frequency bins were used to train separate SVMs in a parallel manner. Therefore, each SVM unit was trained on different EEG bands (e.g., 4-6 Hz and 6-8 Hz) and learned different characteristics of the brain signals. Each SVM unit made its own classification decision. The individual results calculated the final classification result using the majority vote method. To avoid draws, an odd number of SVM units were selected. We call this classifier method as Voting SVM. A similar method was presented in \cite{zhang_eeg_2018}; however, they used their algorithm on Common Spatial Pattern features and utilized Bagging to generate random sub-datasets for each SVM unit. In addition, they omitted to use any artifact rejection algorithm.

\subsubsection{Epoch based cross-validation}
N-fold cross-validation was used to test the validity of the feature extraction algorithms with the SVM classifiers. 

First, the data was split into N distinct parts. In each iteration, N-1 parts were dedicated to the train set and one to the test set. From the classifiers, new instances were created, trained with the training set, and measured the classification performance on the test set. This method was repeated N times. The results of each iteration were saved, and the final accuracy was calculated as the average of the individual classification results. We set the N to 5 in all experiments.

Splitting the data on the window level in case of overlapping windows, instead of epoch level, would result in higher but invalid accuracy, which evokes the overfitting problem. In this case, windows that originate from the same epoch might go both to the train and test set, which can be a major error with highly overlapping windows because it means that almost equivalent features are put on both the train and test set. Therefore, we split the train and test set data on the epoch level instead of the window level. As a result, windows originate from the same epoch exclusively used in either the train or test set.

\subsection{Subjects and Experimental setup}
For the real-time working BCI System, we required a tetraplegic subject as a pilot. Therefore, we contacted MEREK, the Rehabilitation Centre for Physically Disabled People in Hungary.

\subsubsection{Subjects}
Two subjects were applied (B. and C., both male) having C5 or higher spinal cord lesions. The injury of each pilot was confirmed and classified by a neurologist concerning the International Standards for Neurological Classification of Spinal Cord Injury. This so-called Medical Check was required before any Cybathlon competition by the organizers. 

Pilot B was 44 years old and had an incomplete C5 Neurological level of injury (NLI). His Asia Impairment Scale (AIS) was B. The additional comments of the neurologist were: “Dysesthesia in palms. No muscle function in the non-key muscles either (on neither scale).”

Pilot C was 38 years old and had a complete C4 NLI, with AIS A, by the time of the experiments. The additional comments of the neurologist were: “Paresthesia in palms and foot. No muscle function in the non-key muscles either (on neither scale).” 

Offline and online experiments have been conducted with the pilots to record our dataset, test the BCI system on them and make them control the online game.

\subsubsection{Ethical permit}
This study was carried out following the Declaration of Helsinki and national guidelines, with written informed consent obtained from all subjects. The study was approved by the United Ethical Review Committee for Research in Psychology (EPKEB reference number: 2018-54).

Before each experiment, each subject was informed about the course of the experiment, and a statement of consent was signed.

\subsubsection{Data acquisition}
For EEG data recording, a 64-channeled ActiChamp amplifier system (Brain Products GmbH, Gliching, Germany) was used with an actiCAP EEG cap, following the 10-20 international convention. POz was selected as the reference electrode; 63 electrodes were available for data acquisition. During the experimental preparation, the impedance of the EEG electrodes was measured and kept under $30k\Omega$, and its impedance value was saved with the recorded EEG files. 

Subjects were seated in front of an LG Flatron L204WT-SF 20" Wide LCD monitor at a viewing distance of about 110-130 cm. For the experimental location, we used rooms equipped with Faraday cage shielding and common rooms without electrical shielding. Using the regular, unshielded room, we aimed to create a similar environment as it would be present in the Cybathlon 2020 competition.

The raw EEG signals were recorded with the BrainVision Recorder program (version: 1.22.0001) without additional software or hardware filters.

\subsection{Offline Paradigm and Experiments}
This section presents the offline paradigm used for EEG recording with our pilots. Moreover, it gives the methods of comparing features corresponding to EEG bands and comparing our BCI pipeline with the state-of-the-art EEGNet. These comparisons are conducted to select the most suitable configuration of the BCI system which could be used for real-time game control.

\subsubsection{Two Choice Paradigm}
The so-called Two Choice Paradigm was designed to simplify the execution of the Physionet task since our pilots reported that they had difficulty executing all four-limb imagination during some trial experimental sessions.

Before an experiment, the pilots were asked to avoid blinking, swallowing, clenching, or any movements or facial expression unrelated to the actual task during the task periods and try to execute only the required MI tasks repeatedly while the fixation cross was present. During the rest period, the paradigm control program presented the next task on the screen in written form. Here the pilots were allowed to blink, swallow, and execute any movement to prepare for the next task. Pilots were instructed to perform the motor tasks for 4 seconds and the rest task for 3 seconds. 

\begin{figure}[htp]
  \centering
  \includegraphics[width=\columnwidth]{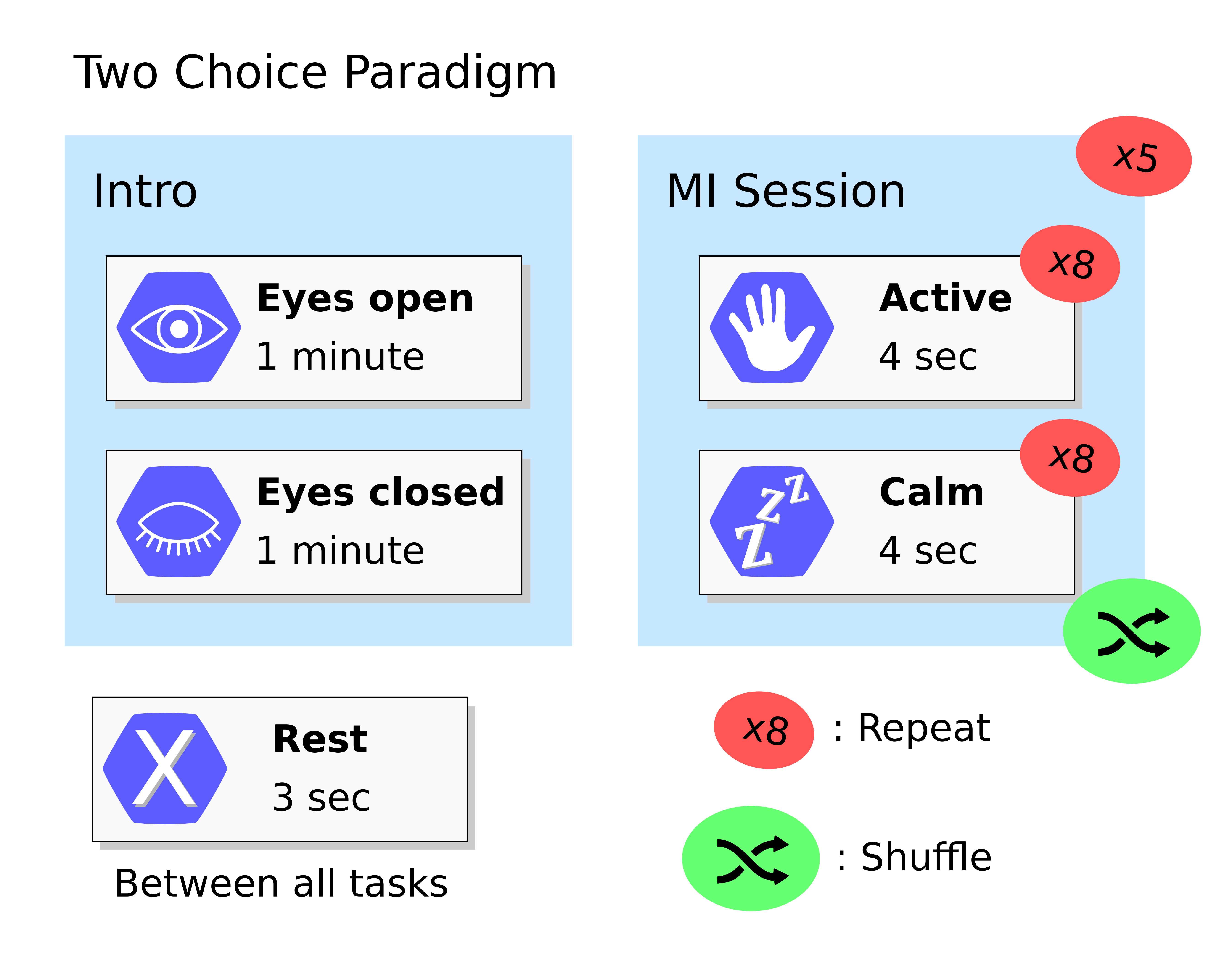}
  \caption{\textbf{Two Choice Paradigm} -- It started with a one-minute open-eye and a one-minute closed-eye task, which served as a baseline and aimed to get the pilots' full attention, preparing them for the MI sessions. Under one MI session, 8 active and 8 calm mental tasks were required from the pilots. The order of the tasks was randomized. The MI session was repeated 5 times under one experiment.}
  \label{fig:paradigm}
\end{figure}

The Two Choice Paradigm, presented in Figure \ref{fig:paradigm}, started with a one-minute-long period when the subject was required to open his eyes and focus on the cross shown on the screen. This session was followed by the instruction of a one-minute-long closed-eye period. In both cases, the subject had to sit as calmly as possible, physically and mentally, without any thoughts. These sessions served as a baseline in the experiments and aimed to get the pilots' full attention, preparing them for the MI sessions.

After the eye sessions, the experiment was followed by 5 MI sessions. Each MI task was presented 8 times in one session, and the order of the task types was randomized. After each completed session, the subjects were allowed to take a break with a self-defined length without leaving the experimental setup.

Under the Active MI tasks, the pilots were allowed to select and combine any motor movements of hand and foot, but it had to be decided and fixed before the start of the experiment. Our tetraplegic pilot B selected left hand and pilot C both leg MI movements for the Active task. The Calm task referred to a task where the subject had to sit with eyes open and omit any movements, thoughts, and possibly other artifacts.

\subsubsection{Investigating the effects of EEG bands on the accuracy}
In order to determine the optimal EEG bands for BCI control, the following experiment was conducted: Several distinct signal processing and classification steps were made, where one classifier received one of the investigated EEG bands, namely: alpha, beta, gamma, theta, range30, and range40. 

The experiments were conducted on both databases. In the case of the Physionet dataset, we made a 4-class classification selecting the active MI tasks: Left Hand, Right Hand, Both Hands, and Both Legs. With our Two Choice Paradigm, a 2-class classification was made. Active imagination of a MI was classified against the calm phase. The active MI task was discussed and fixed with the pilots before the first experiment. Pilot B selected the left hand, and pilot C selected both legs for the Active MI task.

Classification results of different EEG bands were collected and compared with each other statistically. Repeated-measurement ANOVA (rm-ANOVA) was conducted for normally distributed data, followed by two-sided t-tests as a post hoc test. If the data was not normally distributed, the Friedman test was used instead of rm-ANOVA, and the two-sided Wilcoxon signed-rank test was used as a post hoc test. Finally, the p-values were corrected with Bonferroni correction. The significance level was preset to 0.05.

\subsubsection{Comparison with EEGNet}
EEGNet (Lawhern et al., 2018) was used as a standard to evaluate the correctness and precision of our feature extraction and classification algorithms. It was trained with 1-second-long EEG windows generated after the FASTER artifact detection. The training epoch number of the network was set to 500. In order to avoid overfitting, the Early Stopping strategy was used with the patience parameter set to 20. Moreover, the best network weights were saved and restored with a custom strategy before testing: The weights of the network were saved if the validation accuracy was higher or equal and the corresponding validation loss was lower than the previous one. 

The comparison of our method with the state-of-the-art EEGNet was conducted on the Two Choice Paradigm and Physionet database. The normality of the accuracy results was investigated. To determine the significant difference between the methods, a t-test or Wilcoxon test was used concerning the result of the normality test.

\subsection{Online Paradigm and Experiments}
After the offline comparisons of the signal processing and classification algorithms, the most suitable configuration was selected for the real-time BCI system. An online paradigm was created to record data for tuning the classifier of the BCI system and to control the game.

\subsubsection{Online Paradigm} \label{sec:online}
The Online Paradigm was constructed to match the needs in the BCI race of the Cybathlon 2020 competition for tetraplegic subjects (\url{https://cybathlon.ethz.ch/en/event/disciplines/bci}). It started with an offline training period used as calibration for the online game-playing phase. Under the offline training, the Two Choice Paradigm was conducted. The recorded data was used to train the classifier of the BCI system. After the calibration, the BCI system was ready to control the BrainDriver program, which the organizers of the Cybathlon 2020 competition provided. This program handled the BCI discipline's virtual environment and race conditions. A computer monitor displayed the game for a pilot. During the BCI race, pilots got immediate visual feedback from the monitor by observing the result of their (correctly or incorrectly) translated mental commands. 

The BrainDriver program required 4 input commands (3 active commands plus the absence of any commands) from the user, but the Two Choice Paradigm was designed to evoke only two. To overcome this gap, a unique mechanism was introduced, called the Toggle Switch, which was inspired by the Brain Tweaker team (Perdikis et al., 2018). If the user made an active MI task, the game control commands were circulated one after the other with a predefined frequency. When the required control command was reached, the user had to initiate the Calm mental task to keep up the desired command and send no commands at all to the game. The mechanism is presented in Figure \ref{fig:control}.

Using the Online Paradigm, 16 experiments were conducted with the two tetraplegic pilots.

\subsection{BrainDriver game}
The BrainDriver  software is a car racing-like video game developed for the BCI Race of the Cybathlon 2020 in cooperation with ETH Zurich and Insert Coin, Switzerland (\url{http://www.insert-coin.ch/}). A maximum of 4 players (pilots) can compete in this game. Each player must control their avatar, which goes forward by default and reaches the finish line eventually. The goal is to guide these avatars through the racetrack by giving properly timed mental commands in dedicated zones. If an incorrect command is given or no input is given when one is required, the pilot's avatar slows down. On the other hand, giving the proper control command would restore the default speed of the vehicle. 

In this game, there are four types of track elements. Left turn, Right turn, Light on, and Straight zone. In the Light zone, the pilot has to turn on the front light of the vehicle when the surrounding lights go off. On the Straight zone, every command causes the slowdown of the avatar. The Cybathlon organizers fixed the length of the game track to a virtual 500 m, containing 4 pieces from each type of track element.

The BrainDriver can be controlled through the UDP network communication protocol. Each player can send its control command as an unsigned byte code to the server’s IP and port address.

A track generator code was developed for the game, which randomized the order of the different track elements so that a straight track element was followed after each turn. This program was used before each experiment of the Online Paradigm to generate a new game track and prevent the pilots from learning the path.

\subsubsection{Real-time BCI system}
Our real-time working BCI system required prerecorded training data, recorded right before the pilots played with the BrainDriver. Therefore, for signal acquisition, the Online Paradigm was used to create this dataset. 

RBF kernelled Voting SVM was used as a classifier. The classifier of the BCI system was trained offline on the Range40 method with L2 normalization, which was generated from 1-second-long windows with 0.1-second window shifts. 

\begin{figure}[htp]
  \centering
  \includegraphics[width=\columnwidth]{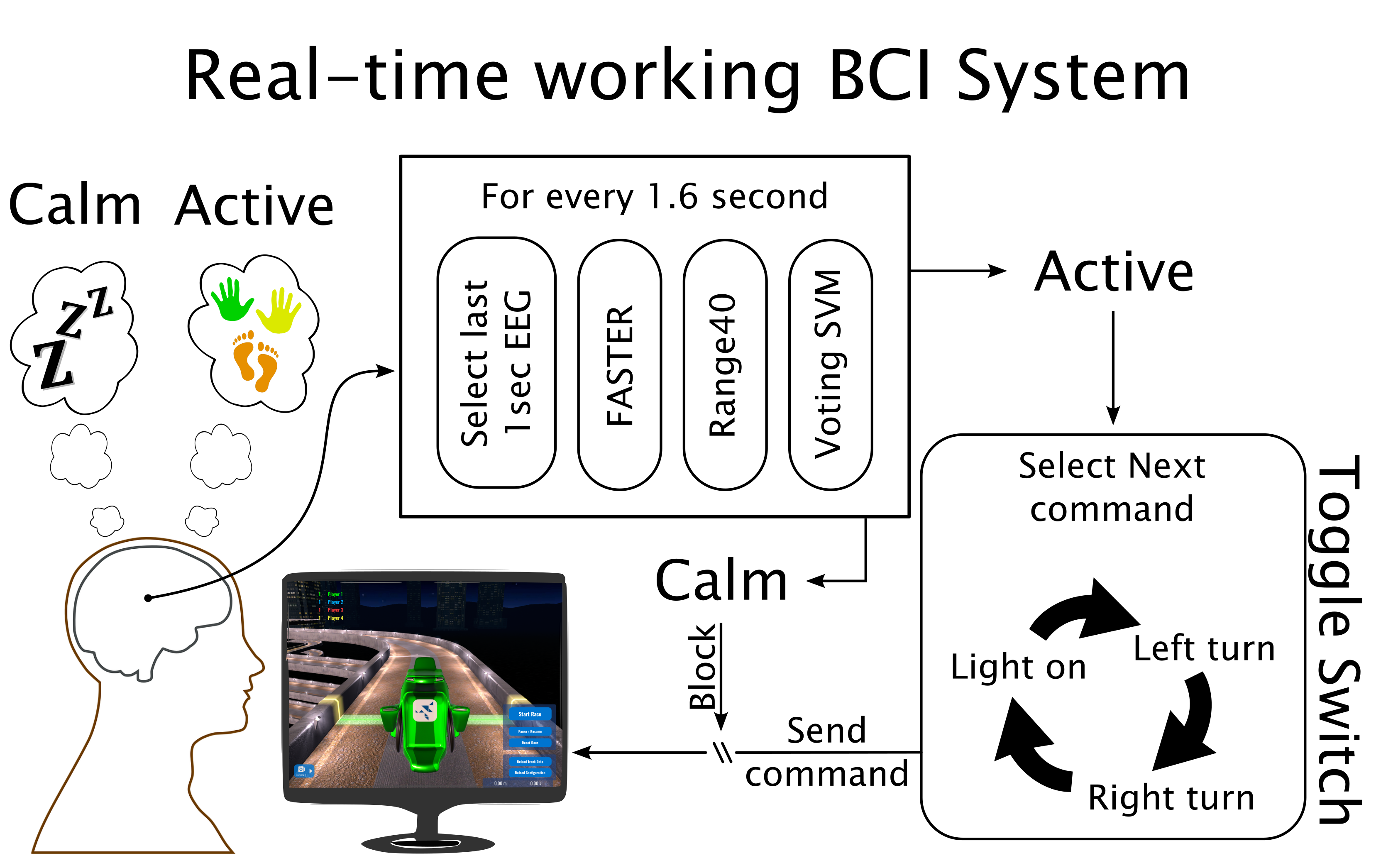}
  \caption{Components of our real-time BCI System and the Toggle Switch control mechanism.}
  \label{fig:control}
\end{figure}

The EEG amplifier received real-time data through the Lab Streaming Layer (LSL) protocol \cite{stenner_sccnliblsl_2021}. Based on the feedback of our pilots, one signal processing and decision-making step was made every 1.6 second, which is the periodic time of sending out a game control command. Under one decision-making step, the latest 1-second-long EEG signals were handled as an EEG window, and the same signal processing and classification steps were used on it as on the prerecorded dataset: the Range40 method was calculated with L2 normalization and sent to the trained RBF kernelled Voting SVM, which classified the signal. The classification result was directly connected with a game control command (Figure \ref{fig:control}) which was immediately sent out to the game server IP address and port number through UDP protocol. The implemented signal processing and classification methods were fast enough to use in the real-time environment. 

Experiments with the real-time BCI system were conducted in a common room of the pilots’ institution, without any electrical shielding to simulate a Cybathlon-like environment.

\section{Results}
\begin{figure*}[htb]
	\centering
    \includegraphics[width=\textwidth]{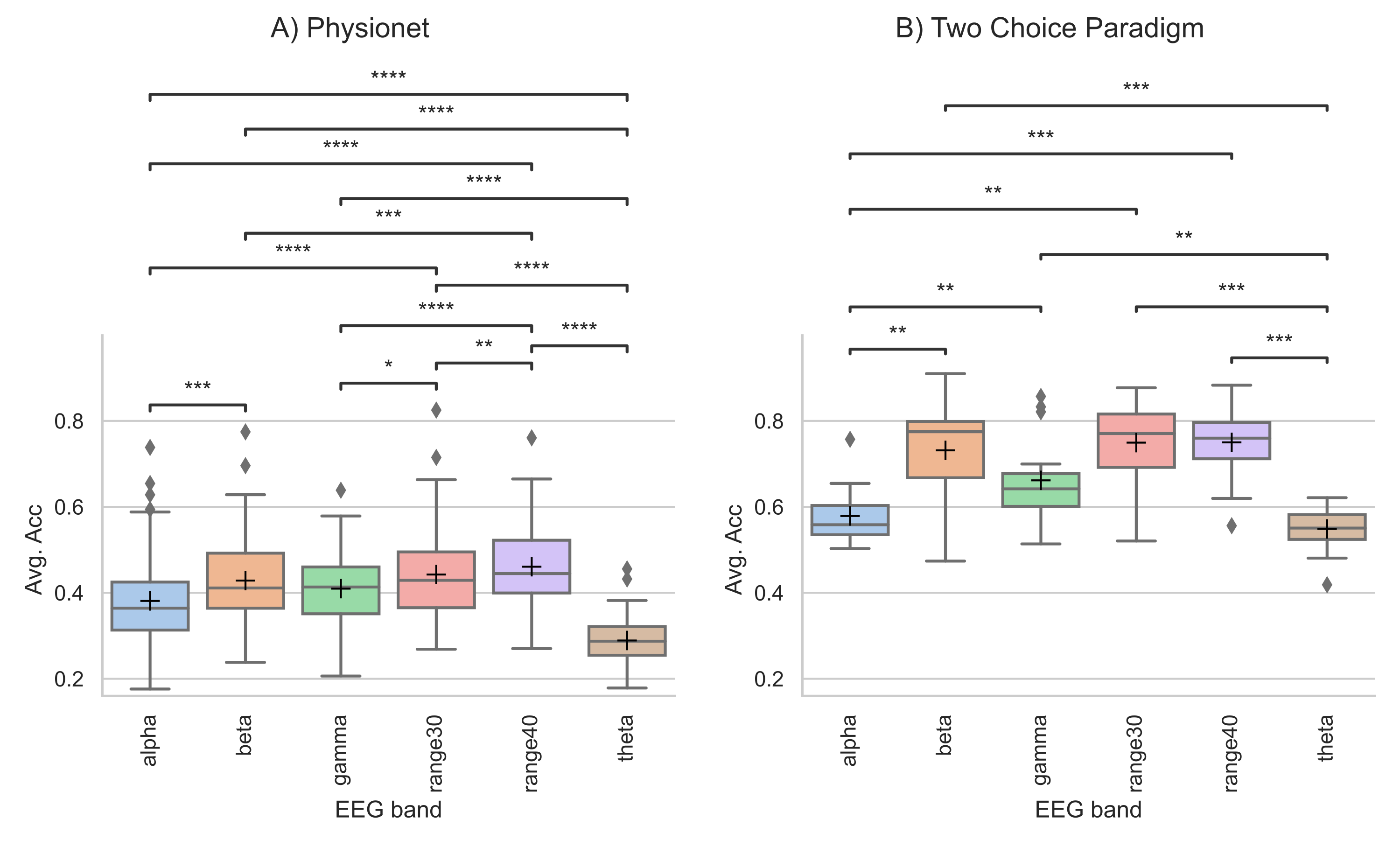}
	\caption{\textbf{EEG band effect investigation} -- On both Physionet and the Two Choice Paradigm database, the impact of different frequency range based features were investigated and compared with each other statistically. The significant differences between the canonical EEG bands and the range30 and range40 methods are marked with stars. The p-value annotation legend is the following: non-significant: 5e-2 < p; *: 1e-2 < p <= 5e-2; **: 1e-3 < p <= 1e-2; ***: 1e-4 < p <= 1e-3; ****: p <= 1e-4. The mean of the data is presented with the '+' symbol.}
	\label{fig:eeg_band_inv}
\end{figure*}

In this chapter, we present our findings concerning EEG band investigation on both Physionet and our own Two Choice Paradigm database, recorded with the help of the pilots. We also show the results of comparing our Voting SVM classifier on the range40 feature with the state-of-the-art EEGNet. Moreover, we present the results made with the real-time BCI application, where we measured the time required from the pilots to reach the finish line of the BrainDriver game.

\subsection{Investigating the effect of EEG bands to the Accuracy}

\begin{table}[htp]
\centering
\caption{Statistical test results of main effect on EEG band investigation}
\begin{tabular}{cccc}
\textbf{Database}       & \begin{tabular}[c]{@{}c@{}}\textbf{Normal}\\ \textbf{distribution}\end{tabular} & \textbf{Stat. test} & \textbf{p-value}   \\ \hline
Physionet       & False                                                         & Friedman   & \textit{\textbf{1.612e-45}} \\
Two Choice Par. & False                                                         & Friedman   & \textit{\textbf{1.533e-09}}
\end{tabular}
\label{tb:main_effect}
\end{table} 

In the case of the Physionet database, a 4-class classification, while in the Two Choice Paradigm, a 2-class classification was made. The classification’s accuracy level was measured for each experiment, which was obtained after a 5-fold cross-validation. The final accuracy concerning a database was determined as the average of all the 5-fold cross-validated experimental accuracies for all subjects (Avg. Acc). These results are presented in Figure~\ref{fig:eeg_band_inv}. Statistical tests were used to determine significant differences between the EEG bands: First, normality tests were used on the 5-fold cross-validated accuracies to determine the type of statistical tests that can be used. 

Table \ref{tb:main_effect}.\ presents the results of the Normality tests, the type of the main effect statistical tests, and the corresponding p-values. The 5-fold cross-validated results did not follow a normal distribution; therefore Friedmann test was used on both Physionet and Two Choice Paradigm databases. The main effect was significant on both datasets; therefore, the Wilcoxon test was used to determine the EEG band, which significantly can produce higher accuracies on the classification side. 

As we can see in Figure \ref{fig:eeg_band_inv}A on Physionet, the beta band from the canonical EEG bands achieved the highest accuracy with 0.4285. It significantly outperformed all the canonical EEG bands but the gamma, which reached a 0.4097 accuracy. However, if we include the range methods, range40 significantly reaches the highest accuracy, 0.4607. 

On the Two Choice Paradigm, presented in Figure~\ref{fig:eeg_band_inv}B, we received similar results as in the case of the Physionet; however, these were less significant. The accuracy of beta, range30, and range40 were 0.7314, 0.7494, and 0.75, respectively. There was no significant difference between these EEG bands; however, this dataset contained 16 experiments compared to the Physionet, which had 105 after exclusion.

\subsection{Comparision with EEGNet}
\begin{figure}[tp]
  \centering
  \includegraphics[width=\columnwidth]{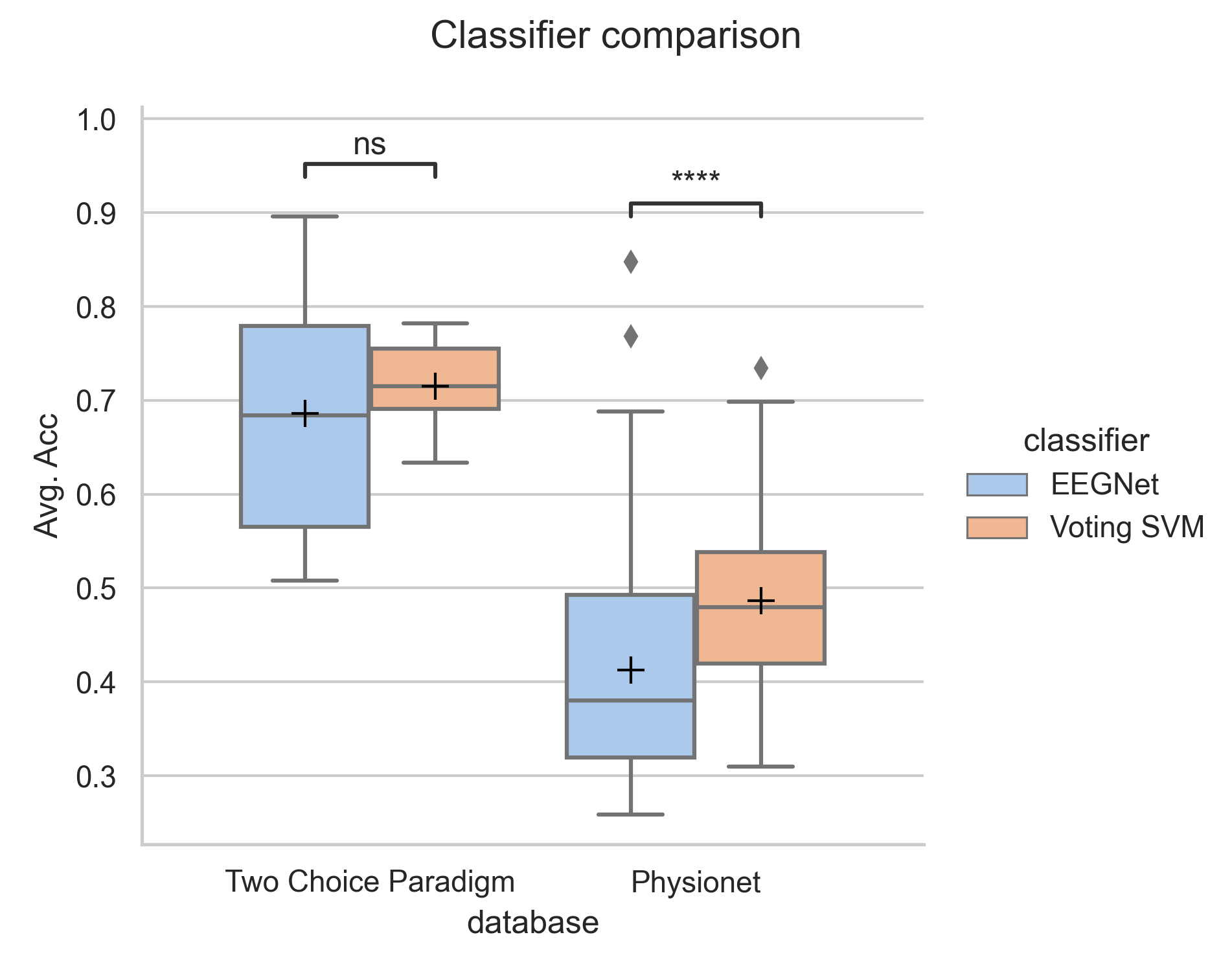}
  \caption{5-fold cross-validated accuracy level comparison of range40 + Voting SVM with EEGNet. The p-value annotation legend is the following: non-significant: 5e-2 < p; *: 1e-2 < p <= 5e-2; **: 1e-3 < p <= 1e-2; ***: 1e-4 < p <= 1e-3; ****: p <= 1e-4. The mean of the data is presented with the '+' symbol.}
  \label{fig:eegnet_comp}
\end{figure}

The Voting SVM classifier with the range40 method was compared with the state-of-the-art EEGNet. On the Two Choice Paradigm, they reached accuracy levels of 0.7151 and 0.6857, while on the Physionet, 0.4866 and 0.4126, respectively. The EEGNet on the Physionet database showed a non-normal distribution of the results; therefore, the Wilcoxon significance test was used. The significance level was preset to 0.05. On Physionet, our Voting SVM with the range40 feature significantly outperformed the EEGNet (p-value < $10^{-4}$). On the other hand, the difference between the two methods was insignificant in our Two Choice Paradigm. The results of the comparisons are presented in Figure \ref{fig:eegnet_comp}.

\subsection{Real-time working BCI Experiment}
\begin{figure*}[b]
  \centering
  \includegraphics[width=\textwidth]{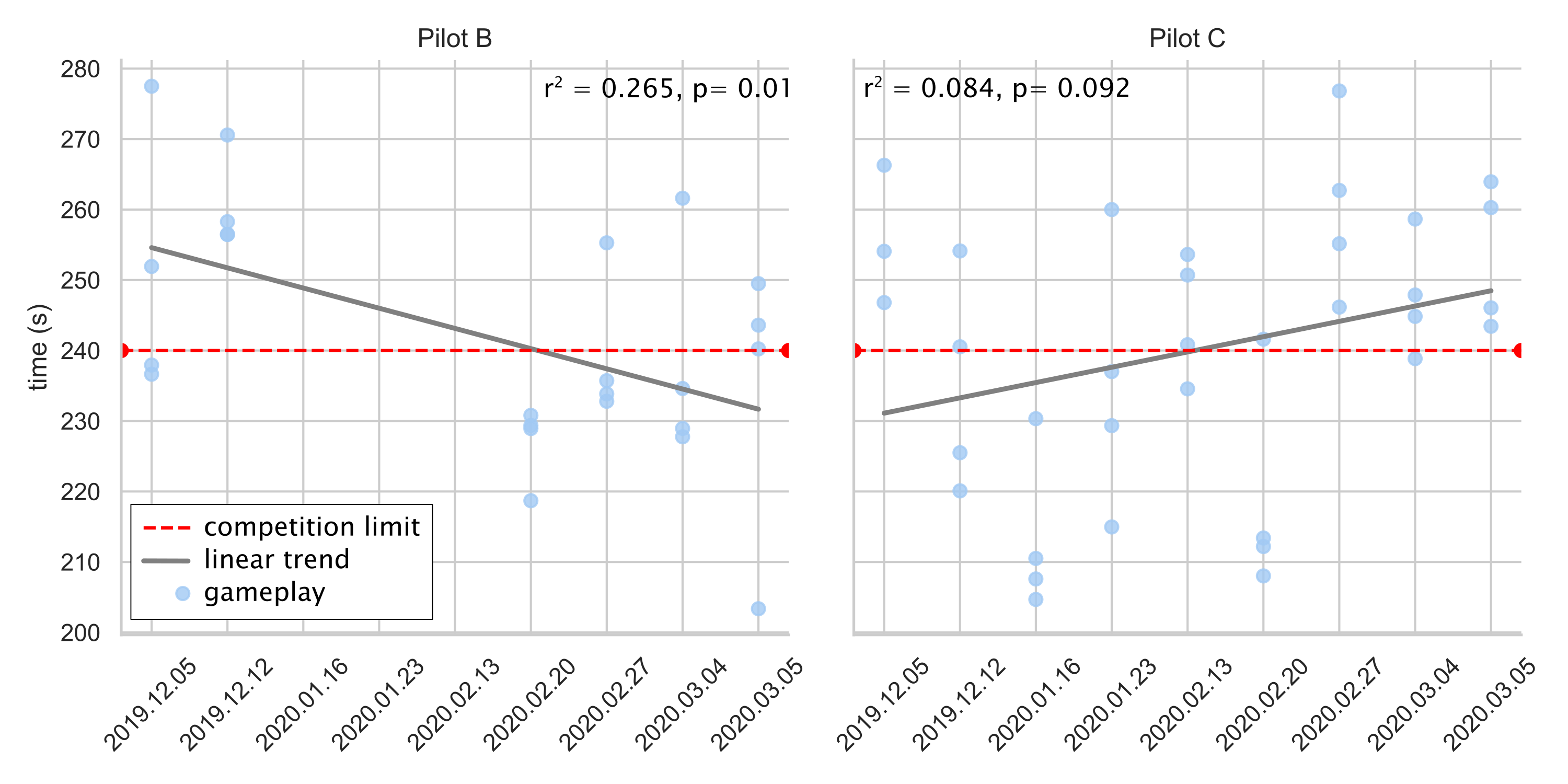}
  \caption{Gameplay performance of pilots per experimental day. 240 seconds were marked with a red line, which is the time limit defined by the organizers. The gray lines present the learning curves.}
  \label{fig:braindriver-gameruns}
\end{figure*}

Using the online Paradigm, we made 59 gameplay sessions in total on 9 experimental days with the help of pilots B and C. The experiments were conducted in a common room of the pilot’s institution, without any electrical shielding. 

Each pilot played with the BrainDriver four times on an experimental day. For each gameplay, we measured the time required to reach the finish line, presented in Figure \ref{fig:braindriver-gameruns}. 240 seconds are highlighted with red because this was the time limitation that the organizers of the Cybathlon 2020 set. 27 gameplays were under this time limit. We also present the learning curves of each pilot in Figure \ref{fig:braindriver-gameruns}. Pilot B showed significant improvement (p-value 0.01 < 0.05); however, it is important to note that he was absent from 3 experiments. On the other hand, Pilot C showed an insignificant increasing learning curve (p-value 0.092 > 0.05).

\section{Discussion}
In this article, we presented the development path of our BCI system, which was created for the Cybathlon 2020 competition. We selected the Physionet \cite{goldberger_ary_l_physiobank_2000, schalk_bci2000_2004} database to test our code instead of the BCI competition datasets because this dataset contains experimental data from 109 subjects, while the others have less than 10. Using the Physionet dataset, we got reliable information about the performance of our algorithm, and significance comparisons were conducted.

Next to the Physionet database, we created our dataset for the real-time BCI system with the help of tetraplegic subjects, called pilots, with C5 or higher spinal cord lesions. We designed a Two Choice Paradigm instead of a standard 4 choice, as Physionet, because the pilots reported having difficulty executing all four-limb imagination.

Focusing on the BCI system, we implemented the FASTER algorithm \cite{nolan_faster_2010} because removing artifacts was a high requirement from the Cybathlon organizers. 

Our BCI system works within the frequency domain. The absolute of the FFT spectrum was calculated for 1-second-long EEG windows as a feature. From the FFTabs, we either calculated the average between two frequencies or many averages from 2 Hz wide, non-overlapping frequency bins, called the Range method. In the case of the Range method, multiple SVMs were trained, where each SVM received only one of the frequency bins. The final decision was determined as the maximum vote of all SVM units. We call this ensemble classifier VotingSVM. To the best of our knowledge, the VotingSVM combined with the range40 method, based on FFTabs, were not investigated and compared statistically on MI databases, nor were used for controlling a computer game as part of a BCI application. We found a similar in \cite{zhang_eeg_2018}; however, they used their algorithm on Common Spatial Pattern features and utilized Bagging to generate random sub-datasets for each SVM unit. The MI tasks were the left little finger and tongue. In addition, they omitted to use any artifact rejection algorithm before the signal processing step. Their proposed algorithm was not compared with any other published signal processing and classification methods.

We conducted multiple comparison analyses on the Physionet and our database to find the most suitable configuration for our BCI system. First, different EEG bands and the range30 and range40 methods were compared statistically. On the Physionet dataset from the canonical EEG bands (alpha, beta, gamma, theta), the beta band reached the highest accuracy level, and the differences were all significant but compared with the gamma band. However, adding the range methods to the EEG bands, the range40 significantly outperforms the rest. Therefore, the range40 was selected with the VotingSVM classifier to compare it with the state-of-the-art EEGNet \cite{lawhern_eegnet_2018} algorithm to correctly highlight and interpret our findings from a broader perspective of the BCI community. According to the Wilcoxon statistical test, our method significantly outperforms the EEGNet. Redoing these tests on the Two Choice Paradigm concluded in less significant results. 

The results of EEGNet were initially presented on the BCI Competition IV 2a database  \cite{tangermann_review_2012}, where it reached an accuracy level of 0.6547 in the case of 4-class classification on 9 subjects. On the Physionet database, we got an accuracy level of 0.4126 with 4-class classification on 105 subjects. We assume these two analyses are not fully comparable because the Physionet database contains 11 times more subjects than the BCI Competition IV 2a dataset. In addition, we included the FASTER algorithm to filter artifacts from the source signals and forced the classifiers to learn on pure EEG signals. To get statistically significant results about differences between classifiers, we suggest using databases with a high number of subjects. Our Two Choice Paradigm based dataset contains data from 16 experimental sessions from 2 disabled subjects. We assume that this is the cause of the statistically insignificant difference between the EEGNet and the Voting SVM on this database.

We created the real-time BCI system after the comparisons. This system contains a unique control protocol called the Toggle Switch. With this algorithm's aid, our pilots could control the BrainDriver computer game with only 2 mental commands instead of 4. This code was inspired by Perdikis et al. \cite{perdikis_cybathlon_2018}, who created an algorithm that classified two MI signals with a thresholding technique. When the third active game control command was required, the pilot initiated the two different active MI tasks in a given time window. In comparison, our method circulates the active control commands one after the other when our pilots initiate the active MI task. Therefore, our approach can easily be extended with additional commands.  

With the Online Paradigm and our BCI system, we conducted real-time BCI experiments with our pilots using the BrainDriver game, developed for the BCI discipline of the Cybathlon 2020 competition. During these gameplays, the pilots received immediate feedback from the computer about the correctness of the given mental commands. Our pilots completed the game with varying runtimes between 200 and 280 seconds, which is reported in Figure \ref{fig:braindriver-gameruns}. Pilot B showed a significant learning curve. However, we would like to highlight that due to the pandemic situation, we could only organize 9 experimental days, resulting in 59 gameplay trials concerning both pilots. 

In the following, we present results from other Cybathlon teams, who participated either in the Cybathlon BCI series 2019 or in the Cybathlon 2020 event, to put our work into perspective.

The Nitro1 team \cite{benaroch_long-term_2021} focused on minimizing the within- and between-session variability and shifts with the aid of the Riemann framework. They projected the generated features to a common reference before making a classification with a Minimum Distance to Mean classifier. A 4-class BCI system was created with 2 MI classes (Left Hand, Right Hand), mental subtraction, and idle state. A blinking detector and thresholding technique on the absolute of the EEG signals were used to reject features with artifacts. Their approach showed increasing classification accuracies; however, this was not reflected in their game performance, measuring the time required to reach the finish line. Most of their runs were over 250 seconds, while in our case, most runs were below.

Team SEC FHT \cite{robinson_design_2021} implemented an artifact removal algorithm, which detected EOG artifacts with Pearson’s correlation and interpolated the affected channels. Filter Bank Common Spatial Pattern was calculated on the purified EEG data, and Gaussian kernelled Support Vector Machine (SVM) was used to classify the features. 4 MI tasks were used (Left Hand, Right Hand, Both Legs, Rest) to evoke control signals. The team investigated the precision of the control system concerning the training protocol. In their analysis, they compared the offline arrow based, the offline game-based, and the online gameplay training. They achieved the best performance with training data derived from the online gameplay, where the pilot received immediate feedback about the correctness of the executed commands. However, considering their best method, training the classifier on data from previous gameplays showed similar fluctuation in gameplay results compared to ours. The finish times were between 210 and 310 seconds, like our method.

The most significant BCI improvement was presented by team MIRAGE91 \cite{hehenberger_long-term_2021}  and WHI \cite{tortora_neural_2022}. In both cases, the regression p-values of the learning curves were under 0.001. The performance range of team MIRAGE91 (160-300 sec) is comparable to our results. On the other hand, team WHI outperformed the rest of the teams. The competition times started from 280 seconds and shrank down to 160. 

Further Cybathlon BCI topics can be found in \cite{korik_competing_2022, muller-putz_editorial_2022, novak_benchmarking_2018, turi_long_2021}.

\section{Conclusion}
This paper introduced an ensemble SVM classifier called VotingSVM, with the range40 feature. To the best of our knowledge, this configuration was not used before in MI-based BCI applications. We critically compared our signal processing and classification algorithm on the Physionet dataset and found that it outperforms the state-of-the-art EEGNet classifier.

Two Choice Paradigm was introduced with the unique Toggle Switch control mechanism in our real-time working BCI system, which was used to control the BrainDriver computer game. 59 gameplay trials were conducted with our 2 tetraplegic pilots, who were both identified with C5 or higher spinal cord lesions. We showed that our results, with respect to the online gameplay, are comparable with the results of other teams who participated in Cybathlon 2020.

We aim to continue the experiments and collect more data to extend our existing dataset, which would be a great opportunity to further develop the BCI system by introducing additional features and normalization techniques and testing neural networks next to the EEGNet. We also would like to focus more on the subject learning because, as Perdikis et al. \cite{perdikis_cybathlon_2018} reported, it significantly impacts the BCI system's robustness. 

As the article shows, our efforts have been encouraging despite the difficulties. Therefore, we are planning to participate in the next Cybathlon event, which will be in 2024.

\section*{CRediT Authorship contribution statement}
\noindent
\textbf{Csaba Köllőd}: Conceptualization, Methodology, Experiments, Software, Statistics, Writing – original draft, Visualization\\
\textbf{András Adolf}: FASTER algorithm implementation, Experiments, Writing - FASTER algorithm\\
\textbf{Gergely Márton}: Experiments, Writing - review\\
\textbf{Moutz Wahdow}: Experiments, Writing - review\\
\textbf{Ward Fadel}: Experiments, Writing - review\\
\textbf{István Ulbert}: Supervision, Project administration, Writing - review \& editing

\section*{Declaration of Competing Interest}
The authors declare that they have no known competing financial interests or personal relationships that could have appeared to influence the work reported in this paper.

\section*{Acknowledgement}
We thank our Pilots for their weekly availability and participation in the experiments. We are also grateful for their regular feedback about the system.

Prepared with the professional support of the Doctoral Student Scholarship Program of the Co-operative Doctoral Program of the Ministry of Innovation and Technology financed from the National Research, Development and Innovation Fund.

\section*{Data availability statement}
The dataset analyzed during the current study is available from the corresponding author upon reasonable request. 

The paradigm leader code can be found at \url{https://github.com/kolcs/GoPar}. For further details about the code, please read Supplementary Materials I. 

Our BCI application’s source code is published at \url{https://github.com/kolcs/bionic_apps}.

\printbibliography
\end{document}